\documentclass[fleqn,usenatbib,twocolumn]{mnras}

\usepackage[english]{babel}
\usepackage{amsmath}
\usepackage{amssymb}
\usepackage{graphicx}
\graphicspath{{./figs/}}
\usepackage{color}
\usepackage{xcolor}
\usepackage{natbib}
\usepackage{mhchem}
\newcommand{\fo}{$f_{\ce{O2}}$}
\newcommand{\ferric}{\ce{Fe^{3+}}}
\newcommand{\ferrous}{\ce{Fe^{2+}}}
\newcommand{\xfer}{Fe$^{3+}/\Sigma$Fe}
\newcommand{\xcore}{Fe$_{\rm mantle}$/(Fe$_{\rm core}$ + Fe$_{\rm mantle}$)}

\title[Exoplanet mantle $f_{{\rm O}_2}$ variability]{A mineralogical reason why all exoplanets cannot be equally oxidising}
\author[C. M. Guimond et al.]{
Claire Marie Guimond,$^{1}$\thanks{Contact e-mail: claire.guimond@physics.ox.ac.uk}\thanks{Present address: University of Oxford, UK}
Oliver Shorttle,$^{1,2}$
Sean Jordan,$^{2}$
John F. Rudge$^{1}$
\\
$^{1}$Department of Earth Sciences, University of Cambridge, Downing Street, Cambridge CB2 3EQ, UK\\
$^{2}$Institute of Astronomy, University of Cambridge, Madingley Road, Cambridge CB3 0HA, UK
}

\date{Accepted XXX. Received YYY; in original form ZZZ}

\pubyear{2023}

\begin{document}
\label{firstpage}
\pagerange{\pageref{firstpage}--\pageref{lastpage}}
\maketitle

\begin{abstract}

From core to atmosphere, the oxidation states of elements in a planet shape its character. Oxygen fugacity ($f_{\rm O_2}$) is one parameter indicating these likely oxidation states. The ongoing search for atmospheres on rocky exoplanets benefits from understanding the plausible variety of their compositions, which depends strongly on their oxidation states---and if derived from interior outgassing, on the $f_{\rm O_2}$ at the top of their silicate mantles. This $f_{\rm O_2}$ must vary across compositionally-diverse exoplanets, but for a given planet its value is unconstrained insofar as it depends on how iron (the dominant multivalent element) is partitioned between its 2+ and 3+ oxidation states. Here we focus on another factor influencing how oxidising a mantle is---a factor modulating $f_{\rm O_2}$ even at fixed Fe$^{3+}$/Fe$^{2+}$---the planet's mineralogy. Only certain minerals (e.g., pyroxenes) incorporate Fe$^{3+}$. Having such minerals in smaller mantle proportions concentrates Fe$^{3+}$, increasing $f_{\rm O_2}$. Mineral proportions change within planets according to pressure, and between planets according to bulk composition. Constrained by observed host star refractory abundances, we calculate a minimum $f_{\rm O_2}$ variability across exoplanet mantles, of at least two orders of magnitude, due to mineralogy alone. This variability is enough to alter by a hundredfold the mixing ratio of SO$_2$ directly outgassed from these mantles. We further predict that planets orbiting high-Mg/Si stars are more likely to outgas detectable amounts of SO$_2$ and H$_2$O; and for low-Mg/Si stars, detectable CH$_4$, all else equal. Even absent predictions of Fe$^{3+}$ budgets, general insights can be obtained into how oxidising an exoplanet's mantle is.

\end{abstract}


\section{Introduction}

Chemical redox equilibria are ubiquitous in all geologic systems. Through the transport of electrons, redox equilibria govern large-scale aspects of the rocky planets built on these systems: from the formation of iron cores \citep{wood_accretion_2006, elkins-tanton_coreless_2008, rubie_accretion_2015, lichtenberg_redox_2021}; to the production of magma \citep[e.g.,][]{holloway_highpressure_1992, foley_reappraisal_2011, stagno_oxidation_2013, lin_oxygen_2021}; the supply of volcanic gas, including greenhouse gases \citep[e.g.,][]{kasting_mantle_1993, delano_redox_2001, gaillard_theoretical_2014, ortenzi_mantle_2020, guimond_low_2021, liggins_growth_2022}; and the availability of chemical species for synthesising biologic precursors and catalysing metabolic reactions \citep[e.g.,][]{muchowska_synthesis_2019, wade_temporal_2021}. Thus the surface environments and atmospheres of rocky planets are profoundly influenced by how oxidising their interiors are. This concept, cemented early on \citep{kasting_mantle_1993, holland_volcanic_2002, kasting_evolution_2003}, is now crucial to bear in mind as we prepare to detect atmospheres on temperate rocky planets, and will hope to distinguish their possible biological origins from geological ones \citep{wordsworth_redox_2018, wogan_abundant_2020, krissansen-totton_oxygen_2021, krissansen-totton_understanding_2022}. The aim of this work is to test whether rocky exoplanets, which are expected to have diverse mantle compositions \citep{hinkel_star_2018, putirka_composition_2019, guimond_mantle_2023, spaargaren_plausible_2022}, will also have more or less oxidising mantles as a consequence.

\subsection{Quantifying how oxidising a planet's mantle is}\label{sec:quantifying-how-oxidising}

The fugacity of a non-ideal gas like oxygen, \ce{O2}, is the effective partial pressure it would have as an ideal gas with the same chemical potential. This seemingly abstract concept is a powerful tool for quantifying how reducing or oxidising a system is. \ce{O2} appears in numerous redox equilibria in the interiors and at the surfaces of planets: both directly, e.g., in the homogeneous gas phase equilibrium \ce{H2 + 1/2O2 = H2O}; and indirectly as metal oxides' variable valence state, e.g,. in the heterogeneous disproportionation equilibrium \ce{3FeO = Fe + Fe2O3} between three valence states of iron \citep{frost_experimental_2004, frost_redox_2008}. It is through the presence of such equilibria that the fugacity of oxygen (\fo) is constrained and can be quantified. This is possible \emph{even when the system has no free \ce{O2} phase}---\fo\,still remains a useful descriptor of how oxidising the system is, by representing the fictive partial pressure of an oxygen gas in equilibrium with the system. 

The processes that influence \fo\,inside a planet are intrinsically linked to the oxidation of Fe, since---as a fact of stellar nucleosynthesis---Fe is by far the most abundant multivalent element found in rocky planets. Therefore, the distribution of Fe between its reduced and oxidised states is interconnected with \fo. We can illustrate this with an inside-out example for planets. 

Hot and partially molten from the energy of accretion, a forming planet's magma ocean might contain Fe co-existing in both its most reduced, metallic, form (Fe$^0$) and its ferrous oxide form (\ce{Fe^{2+}}). The oxidation of iron metal to iron(II) oxide can be written as:
\begin{equation}\label{eq:iw}
\ce{ $\underset{\text{Metallic iron}}{\ce{Fe^0}}$ + \frac{1}{2} O2 <=> $\underset{\text{W\"ustite}}{\ce{Fe^{2+}O}}$ }.
\end{equation}
In our example, the activities, $a$, of Fe-metal and FeO (as dissolved species in the metal and magma phases respectively) would be inter-related with \fo\,through the equilibrium constant of (\ref{eq:iw}):
\begin{equation}\label{eq:K_iw}
K_{(1)} = \exp{\left(\frac{-\Delta G^\circ}{RT}\right)} = \frac{a_{\ce{FeO}}}{a_{\ce{Fe}}\left(f_{\ce{O2}}\right)^{1/2}},
\end{equation}
where $\Delta G^\circ$ is the Gibbs free energy of reaction (\ref{eq:iw}) in ${\rm J}\,{\rm mol}^{-1}$, $R$ is the gas constant in ${\rm J}\,{\rm mol}^{-1}\,{\rm K}^{-1}$, and $T$ is temperature in K. From the form of (\ref{eq:K_iw}), it can be seen that for a system with magma and metal in equilibrium, this system will have higher \fo---be more oxidising---when there is proportionally more FeO in the magma (assuming for simplicity that the activity of FeO scales with its concentration).

For pure phases of Fe-metal and FeO in equilibrium, the activities in (\ref{eq:K_iw}) would always be unity, offering just one possible \fo\,for a given temperature and pressure. Hence, the known \fo\,of equilibria such as (\ref{eq:iw})---the iron-w\"ustite (IW) buffer---are often used as reference values, wherein the \fo\,of a real system is reported as a difference from what this reference \fo\,would be at the temperature and pressure of interest.  %(note this relative \fo\,is only preserved across small $T$, $P$ ranges). 

In a natural system, estimating \fo\,requires knowing the activity of Fe-bearing components in each phase. Magma oceans are highly unlikely to be pure Fe metal and FeO in nature, and calculating these activities and thus absolute \fo, especially at high temperature and pressure, is non-trivial \citep[e.g.,][]{righter_redox_2012}. We can see how this complication comes into the calculation of \fo\,by rewriting the activities in terms of the concentration of the species in the phase, $X_i$, and the rational activity coefficient, $\gamma_i$, which captures departure from ideality. For FeO in a magma from equilibrium \eqref{eq:iw}:
\begin{equation}\label{eq:a_FeO}
{a^\text{magma}_\text{FeO} = \gamma^\text{magma}_\text{FeO}X_\text{FeO}}.
\end{equation}
In this case, whilst $X_\text{FeO}$ would be fixed by the overall amount of oxygen available, the rational activity coefficient may vary as a function of pressure, temperature, and the wider composition of the silicate melt.  As a consequence, \fo\, in this scenario can change \emph{even when the amount of oxygen is constant}; i.e., when ${X_\textrm{Fe}/X_\textrm{FeO}=\textrm{constant}}$.  

The above example makes the point that a complex natural system can become more or less oxidising independently of the amount of oxygen in it. This is true for solid systems of mineral equilibria, as much as for the liquid-liquid `core-forming' equilibria looked at in reaction \eqref{eq:iw}. It is these effects of activity on \fo\,that we investigate in this study, since they can be predicted independently of knowing the (hard to constrain) oxygen budget of the planet---which, simplistically, we consider here to be the ratio of \ce{Fe^{3+}} to \ce{Fe^{2+}} in a mantle rock.

\subsection{Mantle mineralogy and \fo}

The objective of this study is to calculate the effect of solid phase equilibria on mantle \fo\,for exoplanets of diverse composition. We are particularly concerned with the \fo\,at the top of an exoplanetary mantle: not only are the thermodynamic data better constrained at these lower temperatures and pressures \citep[$\lesssim 25\,$GPa;][]{guimond_mantle_2023}, but communication with the surface environment is more direct here. Magmas will likely be generated preferentially in the shallow mantle (as on Earth), and consequently, inherit its \fo, which will set the speciation of gases supplied to the atmosphere during volcanism \citep{gaillard_redox_2015, ortenzi_mantle_2020, liggins_can_2020, guimond_low_2021}. 

Whereas equilibrium \eqref{eq:iw} sets a mantle's \fo\,during core formation, various subsequent mechanisms will likely move the mantle \fo\,away from this initial value \citep[e.g.,][]{frost_experimental_2004, wade_core_2005, wood_accretion_2006, williams_isotopic_2012, rubie_accretion_2015, hirschmann_magma_2022}. Through these processes, mantles can become oxidised such that iron exists between its 2+ (FeO) and its more oxidised 3+ (\ce{Fe2O3}) valence states, as is the case for Earth's mantle. The relevant equilibria setting how oxidising a mantle is, near its surface and some time after core formation, therefore involve minerals incorporating iron in these two oxidation states.

A useful reaction for illustrating this scenario is an equilibrium between the minerals olivine, spinel, and quartz. Here, iron can be present as FeO in olivine, as the component fayalite, and as a \ce{Fe2O3} component in spinel, called magnetite:
\begin{equation}\label{eq:qfm}
    \ce{$\underset{\text{Fayalite}}{\ce{3 Fe_2^{2+}SiO4}}$ + O2 <=> $\underset{\text{Magnetite}}{\ce{2 Fe^{2+}Fe2^{3+}O4}}$ + $\underset{\text{Quartz}}{\ce{3 SiO2}}$}.
\end{equation}
If olivine and spinel are solid-solution phases in equilibrium \eqref{eq:qfm}, \fo\,would be modulated upwards if either \textit{(i)} the fayalite component in olivine is diluted by olivine's other constituents (e.g., the forsterite component, \ce{Mg2SiO4}); or \textit{(ii)} the magnetite component is concentrated in spinel, independently of the amount of FeO or \ce{Fe2O3}. This is because concentration or dilution changes the activities of these components \eqref{eq:a_FeO}.

Like with the example of iron-w\"ustite reaction above, for the pure end-member phases, equilibrium \eqref{eq:qfm} provides a commonly-used reference \fo, that of the fayalite-magnetite-quartz (FMQ) buffer. FMQ has an \fo\,close to that seen in Earth's upper mantle, which is a coincidence resulting from Earth's mantle's composition (Earth's mantle will not have quartz present).
%This is despite orthopyroxene's chemical formula not normally being written with \ce{Fe^{3+}} as a component. Yet together with Al, \ce{Fe^{3+}} can substitute into orthopyroxene as \ce{MgFe^{3+}AlSiO6} \citep{annersten_ferric_1978}, such that an \ce{Fe^{3+}}-bearing component is dissolved in the solid solution that is orthopyroxene. %The extent of \ce{Fe^{3+}} substitution in orthopyroxene itself is independent of \fo\, \todo{(true?? REF)}.

%This fact in itself leads to a crucial part of upper mantle \fo\,variability across different planets. Consider an increase in the proportion of orthopyroxene to Earth's upper mantle's first-most abundant upper mantle mineral, olivine---whose crystal structure takes in no \ce{Fe^{3+}}---whilst keeping the total ratio \ce{Fe^{3+}} to \ce{Fe^{2+}} constant. The larger orthopyroxene mass for the same \ce{Fe^{3+}} mass means that \ce{Fe^{3+}} is diluted in the orthopyroxene solid solution. Hence the activity of the \ce{Fe^{3+}}-bearing component in orthopyroxene decreases---\ce{Fe^{3+}} is less ``reactive". Analogously to $a_{\ce{FeO}}$ in (\ref{eq:K_iw}), \fo\,also must decrease, all else equal.

Quartz is also not expected to be a common mineral in the shallow mantle mineralogies of many rocky exoplanets \citep{putirka_composition_2019, guimond_mantle_2023, spaargaren_plausible_2022}. Rather, in most planets as on Earth, it is equilibria among olivine, pyroxenes, garnet, and spinel that set the \fo\,\citep[see][for a detailed review of these effects]{stolper_effects_2020}. The fact that these multi-mineral equilibria govern \fo\,means that the effect of shifting phase proportions on \ce{Fe2O3} activities will everywhere modulate the \fo\,of planetary mantles \citep[e.g.,][]{frost_introduction_1991, oneill_ferric_1993, ballhaus_upper_1995, rohrbach_metal_2007, frost_redox_2008, jennings_simple_2015, stolper_effects_2020}. For Earth, this effect has been investigated in the context of an isochemical mantle (i.e., constant abundance ratio between all oxides, including \ce{FeO} and \ce{Fe2O3}) existing at different pressures, where order-of-magnitude changes in \fo\,were predicted just because of pressure-dependent mantle mineralogy \citep{stolper_effects_2020}. 

Beyond the solar system, however, mineral phase proportions \textit{at a given pressure} will vary significantly according to the bulk mantle oxide proportions of a planet. This mineralogical effect on \fo\,can therefore be calculated for unknown exoplanets by modelling their plausible mineral phase equilibria, based on estimates of their bulk mantle metal oxide compositions, which, are expected to reflect the refractory element ratios observed in their host stars \citep{anders_solarsystem_1982, thiabaud_elemental_2015, bonsor_hoststar_2021}. In this way the chemical ``star-planet connection" \citep{hinkel_star_2018} continues to be explored \citep[e.g.,][]{unterborn_thorium_2015, dorn_can_2015, dorn_generalized_2017, dorn_bayesian_2017, santos_constraining_2017, dorn_new_2019, unterborn_effects_2017, unterborn_pressure_2019, putirka_composition_2019, wang_enhanced_2019, wang_detailed_2022, otegi_impact_2020, spaargaren_influence_2020, spaargaren_plausible_2022, guimond_mantle_2023, unterborn_nominal_2023, unterborn_mantle_2022}.

%\todo{planet-hosting stellar chemistry variability is a great way to explore what different planetary mantles can look like in general}

\subsection{This study's constraint on exoplanet mantle \fo}

We cannot claim to know \textit{a priori} the ratios of \ce{Fe^3+} to \ce{Fe^2+} in exoplanet mantles, which would be necessary to predict their absolute \fo. A more tractable endeavor is to fix the \ce{Fe^3+}/\ce{Fe^2+} ratio at a nominal value for all planets, and calculate the \textit{variability} of \fo\,due directly to phase equilibria; i.e., the changes in how \ce{Fe^3+} is distributed between co-existing mineral phases and the impact of this on \fo. We go on to show that this variability is largely independent of the chosen \ce{Fe^3+}/\ce{Fe^2+} ratio. By this method we hence produce an estimate of the minimum \textit{variability} in mantle \fo\,across exoplanets: \fo\,must differ by \textit{at least} this much because of mineralogy alone.

Because this \fo\,constraint is obtained from modelling mantle phase equilibria directly, our approach is complementary to previous work by \citet{ortenzi_mantle_2020} and \citet{liggins_growth_2022}, who explored the possibility of inferring mantle \fo\,from anticipated spectroscopic observations of exoplanet atmospheres---certain patterns of gas species in a volcanic atmosphere would in principle link to how oxidising the mantle source is. As for direct observational prospects, \citet{doyle_oxygen_2019, doyle_where_2020, doyle_new_2023} report measurements of FeO abundances in planetary material at the end of its life, accreted in fragments onto white dwarfs. Such measurements would link to the bulk Fe oxidation state of the original parent body given an \textit{a priori} estimate of how its iron inventory was partitioned between metal and oxides. Accurately retrieving parent body oxygen abundances from polluted white dwarfs remains difficult, however, as the perceived abundances can vary over time as material accretes \citep{brouwers_asynchronous_2023}.

We present a first \textit{a priori} constraint on the variability of mantle \fo\,across exoplanets, stemming directly from refractory abundance distribution in the Hypatia Catalog of nearby stars \citep{hinkel_stellar_2014}. In section \ref{sec:methods-mineralogy}, we outline our method of calculating upper mantle mineralogy and the associated \fo\,from bulk mantle compositions, and in section \ref{sec:methods-bulkcomp}, of converting from stellar refractory element abundances to these bulk mantle compositions. Section \ref{sec:results} presents the resulting distributions and various compositional correlations of \fo. Section \ref{sec:discussion} discusses a consequence for planetary evolution; namely, the speciation of outgassed volatiles. Section \ref{sec:conclusion} concludes the study.

\section{Methods}

\subsection{Phase equilibria and oxygen fugacity}\label{sec:methods-mineralogy}

We use two independent models to calculate phase equilibria at pressures $P \in [1\,,4]\,{\rm GPa}$, and temperature $T = 1373\,{\rm K}$, for a wide sample of hypothetical exoplanet bulk mantle compositions. Although we expect these two models to result in different values of absolute $f_{\ce{O2}}$---due to differences in how they incorporate \ce{Fe^3+} into minerals and treat mineral solid solutions---we will show that the compositional variation in $f_{\ce{O2}}$ is broadly robust to the choice of model. There are three components to each model: \textit{(i)} the codes that solve for the stable mineralogy and mineral compositions; \textit{(ii)} the thermodynamic databases, which contain data on endmember mineral component entropies, molar volumes, etc.; and \textit{(iii)} activity-composition relations, or ``mineral models'', that describe the solid solutions and their thermodynamics. 

First, we use the pMELTS software package and its native thermodynamic database and mineral model \citep{asimow_algorithmic_1998, ghiorso_pmelts_2002}. For numerical stability, all calculations are initialised at superliquidus conditions, $2273\,{\rm K}$. Then the system is cooled along the desired isobar in 5-K increments, until the $T$ of interest is reached. The target temperature was chosen to be well below the solidus of typical rocky planet mantles. However, some mantles are predicted to still contain a small fraction of liquid, which will be having a small effect on the estimated \fo.  We therefore exclude any mantle compositions with a remaining liquid phase in excess of 1 wt.\% at the $T$ of interest. About 50 compositions fail to converge at subsolidus temperatures with pMELTS and no stable phase assemblage can be found. 

We perform a second set of calculations using the \citet{jennings_simple_2015} thermodynamic database and mineral model implemented in the code Perple\_X \citep{connolly_geodynamic_2009}; hereafter, JH-15. These calculations are done in the constrained minimisation mode of Perple\_X, at the $T$ and $P$ of interest.

\subsubsection{Pressures and mineral phases considered}\label{sec:methods-pressures}

Perple\_X and pMELTS calculate the relative proportions of stable phases by minimising the Gibbs free energy for an input wt.\% bulk oxide composition (section \ref{sec:methods-bulkcomp}), over an input $T$ and $P$. We consider cases where the solution phases olivine, orthopyroxene, clinopyroxene, spinel, and garnet comprise $\sim$100\% of the mineralogy between 1--$4\,{\rm GPa}$. We exclude any bulk compositions that stabilise a pure-\ce{SiO2} phase (e.g., quartz, coesite) due to less-well-constrained thermodynamic data (see section \ref{sec:discussion-sio2}). We further exclude the handful of compositions where pMELTS stabilises extraneous phases, such as kyanite or rhombohedral-oxide solution. 

Our choice of pressure endpoints of 1 and 4 GPa excludes crustal phases whilst spanning the important spinel-garnet transition in the upper mantle. No subsolidus phase changes are expected at pressures between 4 and $\sim$10$\,{\rm GPa}$ for rocky exoplanet compositions \citep[e.g.,][]{guimond_mantle_2023}. Hence the mineral assemblages we calculate at $4\,{\rm GPa}$ should be representative of the bulk of the potential magma source region \citep{gaillard_diverse_2021}. Although we generally expect more reducing conditions at the higher pressures of the transition zone and lower mantle \citep{frost_redox_2008}, we do not consider these mantle regions in our study: in part because the thermodyamic data is poorly-constrained, but also because these regions would not supply as much oxidising or reducing power to the planetary surface via magmatism.

We note that the hypothetical planet's mass is not directly relevant for our calculations, given we are fixing $T$, $P$; the only implicit constraint on planet mass therefore is that it be large enough to have mantles reaching $4\,{\rm GPa}$ (for context, Mars, 1/10 the mass of Earth, has a mantle pressure up to $19\,$GPa at its base; \citealt{stahler_seismic_2021}).  The direct effect of planet mass in this framework is therefore just to change the depth in km corresponding to the pressure of interest. In practice, planet mass may have influenced the oxygen budget of the mantle \citep[e.g.,][]{frost_redox_2008, deng_magma_2020}, but such effects are not considered in this study.

\subsubsection{Absolute oxygen fugacity from chemical potentials}

The fugacity of \ce{O2} is related to its chemical potential, $\mu$, at the $T$ and $P$ of interest, relative to the standard state chemical potential, $\mu_0$, at 1 bar and the $T$ of interest:
\begin{equation}\label{eq:mu_fo2}
\log_{10} f_{\ce{O2}}= \frac{\mu - \mu_0} { R T \ln(10) },
\end{equation}
where $R = 8.3145\,{\rm J\, mol}^{-1}\,{\rm K}^{-1}$ is the gas constant, and $T$ is temperature in K. Thus by adopting a fixed $T$ and $P$, we can perform meaningful comparisons of \fo\,between different bulk compositions.  

The pMELTS software internally calculates $\log_{10} f_{\ce{O2}}$ at each $T$ and $P$ of interest \citep[see][]{asimow_algorithmic_1998}. Meanwhile, Perple\_X only returns $\mu$, so here we explicitly use (\ref{eq:mu_fo2})\,to find $\log_{10} f_{\ce{O2}}$, with $\mu_0$ calculated via the same JH-15 database.

\subsubsection{Relative oxygen fugacity using buffers}

Because oxygen fugacities strongly depend on $T$ and (less so) $P$, it is convenient to report them as a difference in dex with respect to the value of $\log_{10} f_{\ce{O2}}$ produced by a known buffer reaction---such as (\ref{eq:iw}) or (\ref{eq:qfm})---at the same $T$ and $P$ of interest. For relatively modest changes in $T$ and $P$, intrinsic $\log_{10} f_{\ce{O2}}$ differences between system and buffer will be approximately preserved, thus largely normalising out the direct effect of temperature on $\log_{10} f_{\ce{O2}}$. 
%Because buffers tend to follow roughly parallel $\log_{10} f_{\ce{O2}}$ paths through $P$-$T$ space (within a range). Therefore the $T$ and $P$ information can often be elided.\footnote{See \citet{righter_redox_2012} for discussion of when this approximation breaks down.}
Hence we will also report most of our calculations as a relative $\log_{10} f_{\ce{O2}}$ with respect to FMQ (\ref{eq:qfm}), denoted $\Delta$FMQ. 

We calculate $\log_{10} f_{\ce{O2}}$ of FMQ using JH-15, and subtract the absolute $\log_{10} f_{\ce{O2}}$ from pMELTS or JH-15 to find $\Delta$FMQ. We emphasise, however, that the $\Delta$FMQ values we find are not the key result here, rather their distribution.

\subsection{Mantle bulk composition from stellar element abundances}\label{sec:methods-bulkcomp}

Our method of converting stellar abundances to phase proportions is essentially the same as in \citet{guimond_mantle_2023} and discussed there in more detail. 

We take the entire sample of planet-hosting FGKM stars from the Hypatia Catalog \citep{hinkel_stellar_2014} which have measured Mg, Si, Fe, Ca, Al, Na, and Ti, using the reported mean if a star has been measured more than once. Here, the normalised stellar abundance of an element X with respect to hydrogen is reported as:
\begin{equation}\label{eq:nH_star}
    {\rm [X/H]} = \log_{10}(n_{\rm X}/n_{\rm H})_{\star} - \log_{10}(n_{\rm X}/n_{\rm H})_{\sun},
\end{equation}
where $n$ is the number abundance, the subscript $\star$ denotes the value for the star, and the subscript $\sun$ denotes the solar value from \citet{lodders_abundances_2009}. We conserve the relative number abundances of Mg, Si, Fe, Ca, Al, Na, and Ti between the star and the bulk planet.

Within planet interiors, these seven elements together with oxygen will occur as component oxides (e.g., MgO, \ce{SiO2}) of mantle mineral phases, which constitute virtually its entire mass. We place most of the bulk Fe as a metallic planetary core, leaving an unconstrained but minority fraction as Fe oxide in the mantle. That is, we assume that there is enough O available to bond with all of the Mg, Si, Ca, Al, Na, and Ti; we do not track the affinity of O for rocky or volatile/icy material in the protoplanetary disc. Because the extent of Fe oxidation is unknown \textit{a priori}, we use a free parameter $\chi_{\rm Fe}^{\rm mantle}$~=~\xcore\;which defines the molar ratio of FeO$^*$ in mantle oxides to the total bulk planet Fe---here and throughout we use FeO$^*$ to mean the sum of iron oxides regardless of the valence state of iron (2+ or 3+), whilst FeO denotes ferrous iron oxide alone. We initially use a fixed value of $\chi_{\rm Fe}^{\rm mantle} = 0.12$ (that reproduces Earth's core mass fraction for pure, solid Fe), but later test the effects of varying $\chi_{\rm Fe}^{\rm mantle}$ on the distribution of \fo.

% higher p-t core formation can increase mantle FeO content through Si reduction (Siebert et al. 2012; Fischer et al. 2015),

To find the bulk composition of the mantle in weight percent, being the required input to the thermodynamic models, we convert the abundance of each element [X/H] to its wt.\% equivalent, conserving the total moles of Fe as described in \citet[n.b. here we interpret their FeO as FeO*]{guimond_mantle_2023}.

Planetary mantles will only inherit their host star [X/H] if element X is perfectly refractory. Ca, Al, Mg, Si, Fe, and Ti are highly refractory, so we apply our stellar abundance-to-mantle oxide method at face value (but removing some Fe to the core as above). For Na, which is more volatile, as a first guess we approximate a depletion factor: we calculate the molar ratio of Na/Ti in the bulk silicate Earth \citep{workman_major_2005} relative to Na/Ti in the solar photosphere \citep{lodders_abundances_2009}, and scale [Na/H] by this constant planet/star depletion factor of 0.046 across all systems. \citet{spaargaren_plausible_2022}, in a similar attempt to relate stellar to planetary compositions, additionally account for possible Si in the core and for the slightly higher volatility of Mg versus Si, and hence produce slightly different Mg/Si ratios than our model. The effects of secondary processing on mantle Mg/Si ratios do not affect our \fo\,variability results, but are briefly discussed in \ref{sec:discussion-mgsi}.% Whilst we do not include this effect, the compositional trends we present allow the effect of differential Mg/Si volatility to be interpreted in terms of \fo\,change.

The bulk mantle compositions resulting from this method will be approximations of the true composition of a planet; caveats are discussed more extensively in \citet{guimond_mantle_2023}. However, because it allows us to cover a large sample (\textgreater 1000) of observed stellar compositions, our approach can provide a useful estimate of the corresponding breadth of mineralogically-driven \fo\,variation, regardless of the true median, which is in any case beyond our reach. %\todo{also, because we are interested in the range and can't pin down \fo\,anyways, small changes in mineralogy due to partitioning processes we don't capture will not matter for our resulting distributions, they might just shift the whole thing}

\subsubsection{Nominal ferric iron content}

To calculate \fo\,emerging from Fe redox equilibria, we must specify, as thermodynamic components in the system, either FeO$^*$ and \ce{O2} (Perple\_X input), or \ce{FeO} and \ce{Fe2O3} (pMELTS input). Both ways of describing the composition of the system are equivalent, as some amount of nominal \ce{O2} in fact forms ferric iron to create a unique molar ratio of FeO to \ce{Fe2O3} (again,  assuming no other multivalent, redox-active elements are present in the system). From the stoichiometry of the simple redox reaction \ce{4FeO + O2 <=> 2Fe2O3}, we have 2 mol \ce{Fe2O3} for every mol \ce{O2}, and 2 mol \ce{Fe^3+} for every mol \ce{Fe2O3}, so the effective molar abundance of \ce{O2} is: 
\begin{equation}\label{eq:x_ferric}
n_{\ce{O2}} =  \frac{1}{4}\left(\ce{Fe^{3+}}/\Sigma \ce{Fe}\right) n_{\ce{FeO}^*},
\end{equation}
where \xfer\;is the molar ratio of ferric iron to total iron in the (bulk) mantle, and $n_{\ce{FeO}^*}$ is the molar abundance of Fe in mantle oxides as required by stellar element abundances and $\chi_{\rm Fe}^{\rm mantle}$. 

Equivalently, we can find the component mass fractions of \ce{FeO} and \ce{Fe2O3} by simultaneously solving:
\begin{equation}\label{eq:m_fe2o3}
\begin{split}
\ce{Fe^{3+}}/\Sigma \ce{Fe} &= \frac{2m_{\ce{Fe2O3}} / M_{\ce{Fe2O3}}}{m_{\ce{FeO}} / M_{\ce{FeO}} + 2m_{\ce{Fe2O3}} / M_{\ce{Fe2O3}}}\\
m_{\ce{FeO}^*} &= m_{\ce{FeO}} + m_{\ce{Fe2O3}},
\end{split}
\end{equation}
where $m$ denotes the mass in kg and $M$ the molar mass in ${\rm kg}\,{\rm mol}^{-1}$ for the subscripted species.

In this way, a fixed \xfer\;is imposed across all compositions: for Perple\_X runs \ce{O2} is added to make up the required \xfer\;given FeO$^*$ using (\ref{eq:x_ferric}); or, for pMELTS runs, FeO$^*$ is divided into FeO and \ce{Fe2O3} using (\ref{eq:m_fe2o3}). Note that in either case the total mass or number amount drops out once the bulk composition is re-normalised to 100\%. Earth's mantle \xfer\;value of 3\% would be equivalent to having 8.2 wt.\% \ce{FeO} react with 0.027 wt.\% \ce{O2}. 

Note that different choices of \xfer\;will have a minor effect on equilibrium mineralogy. Increasing \xfer\;from 3\% to 10\%, for example, increases the amount of orthopyroxene by a few wt.\% at the expense of clinopyroxene. %, in effect decreasing the ratio of olivine to orthopyroxene from 1.86 to 1.50 \todo{at XX $T,P$}. 
However, the amount of mineralogically-derived variability in mantle \fo\,is roughly preserved across different values of \xfer, as we will show.

\subsubsection{Choice of oxide components}\label{sec:methods-components}

\begin{figure}
\centering
  \includegraphics[width=1\linewidth]{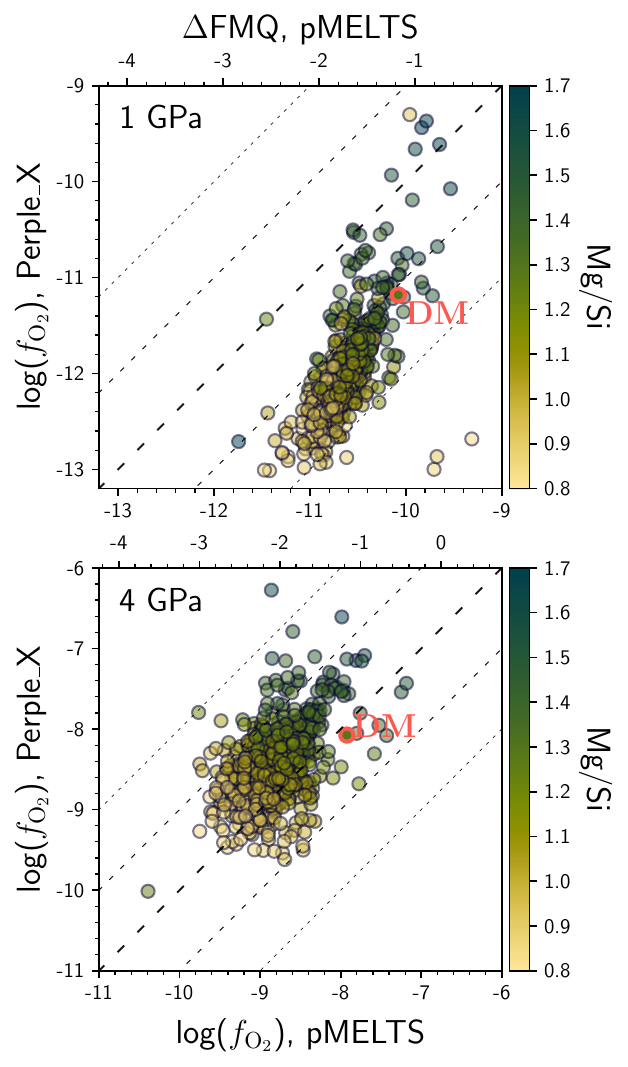}
\caption{\label{fig:model_comp}Direct comparison of absolute \fo\,between the pMELTS and JH-15 in Perple\_X models, at $1\,\text{GPa}$ \textit{(top)} and $4\,\text{GPa}$ \textit{(bottom)} and $1373.15\,\text{K}$. Each point represents the bulk mantle composition inferred from a planet-hosting star in the Hypatia Catalog, assuming ${\rm Fe}^{3+}/\Sigma{\rm Fe} = 3$\% and ${\rm Fe}_{\rm mantle}/({\rm Fe}_{\rm core} + {\rm Fe}_{\rm mantle}) = 12$\%. Earth's depleted mantle (DM) composition from \citet{workman_major_2005} is highlighted in red for comparison. Points are coloured by the molar Mg/Si ratio. Both models' compositions are in terms of MgO, \ce{SiO2}, \ce{Al2O3}, \ce{CaO}, \ce{Na2O}, \ce{Cr2O3}, \ce{FeO}, and \ce{Fe2O3}; pMELTS has \ce{TiO2} in addition.}
\end{figure}

The major rock-forming oxides MgO, \ce{SiO2}, CaO, \ce{Al2O3}, and FeO$^*$ should make up \textgreater~99\% of exoplanetary mantles by mass---considering only these components would suffice if our only goal were to estimate phase abundances. However, other minor oxides can potentially have a non-negligible influence on mantle \fo, either by stabilising important ferric iron-hosting phases at different $T,P$, or affecting \ferric\,substitution in crystal structures. Hence, as in \citet{stolper_effects_2020}, we also consider \ce{Na2O} and \ce{TiO2}, and in one set of reference calculations, \ce{Cr2O3}. 

Although the presence of \ce{Na2O} has minor effects on orthopyroxene phase proportions in both pMELTS and JH-15 (changing them by $\lesssim10$ wt.\%), this can lead to disproportionately large effects on \fo. Including \ce{TiO2} in the pMELTS runs is necessary to ensure that numerically-stable subsolidus phase assemblages can be found for most bulk compositions. Meanwhile, JH-15 can only place \ce{TiO2} in a pure rutile phase, so for Perple\_X runs we exclude \ce{TiO2} from bulk compositions in the thermodynamic modelling, although we still retrieve stellar Ti to calculate the Na depletion as discussed above. Minor amounts of \ce{Cr2O3} affect \ferric\,substitution into pyroxenes, whilst stabilising spinel at slightly higher pressures, allowing Earth's upper mantle \fo\,to be reproduced more accurately from Earth's depleted mantle composition.

\begin{figure*}
\centering
\includegraphics[width=1.05\textwidth]{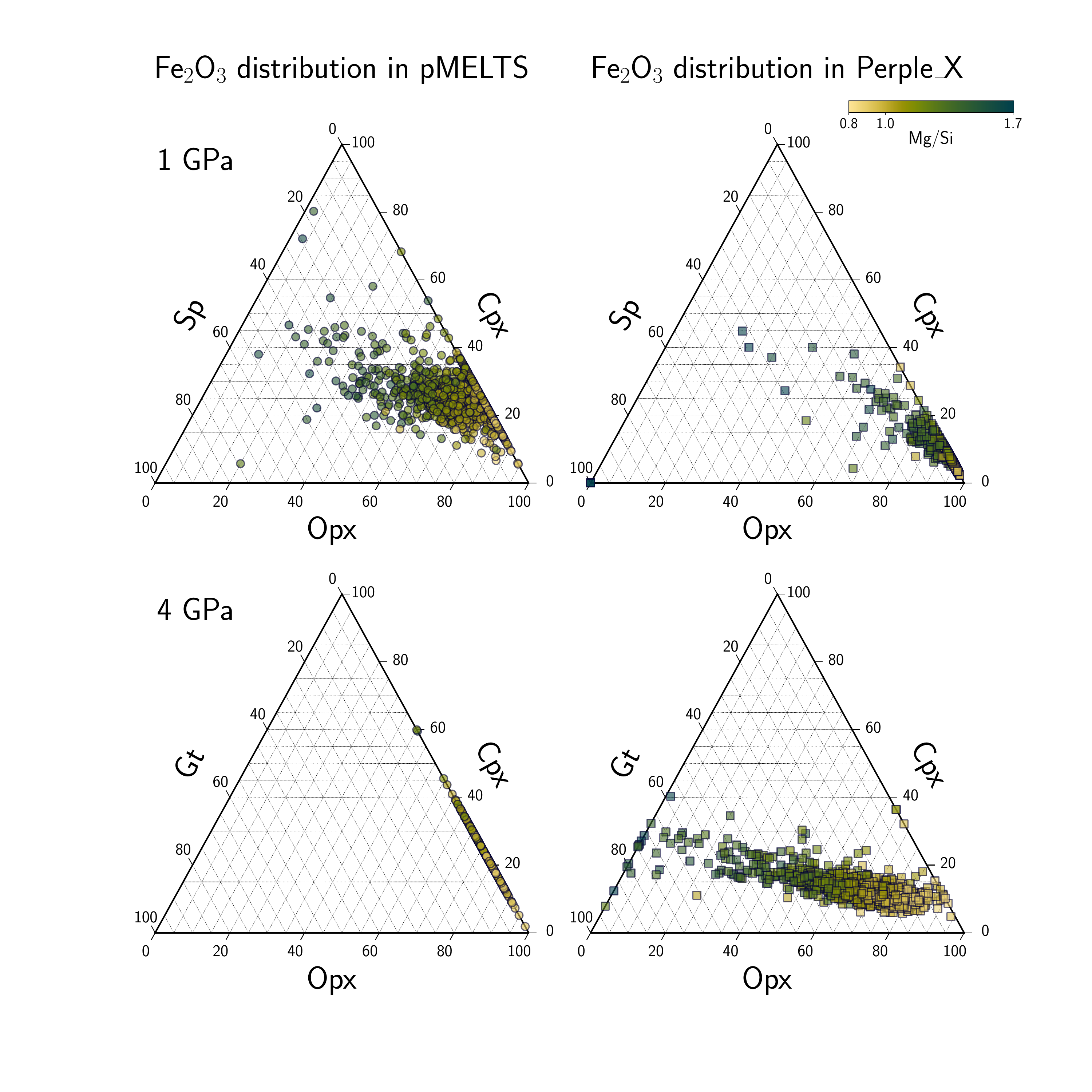}
\caption{\label{fig:ferric_ternary}Ternary diagram showing the distribution (modality) of \ce{Fe2O3} between its subsolidus host phases: orthopyroxene (Opx), clinopyroxene (Cpx), and spinel (Sp) or garnet (Gt), at $1\,{\rm GPa}$ \textit{(top)} and $4\,{\rm GPa}$ \textit{(bottom)} and $1373.15\,\text{K}$. Olivine contains no \ce{Fe^3+} in either of the thermodynamic databases we consider. Fe$_2$O$_3$ modality is calculated as the weight fraction of \ce{Fe2O3} in each phase, multiplied by the phase's total weight fraction in the system, and normalised to 100\% between the three phases on each axis. The right column shows results from pMELTS, and the left column shows Perple\_X results. Each point represents a bulk mantle composition (Ca-Na-Fe-Mg-Al-Si-O, plus Ti in pMELTS) inferred for a planet-hosting star in the Hypatia Catalog. Points are coloured by Mg/Si. Note that the horizontal guides correspond to the Cpx axis.}
\end{figure*}

\section{Results}
\label{sec:results}

\begin{figure*}
\centering
\includegraphics[width=1\textwidth]{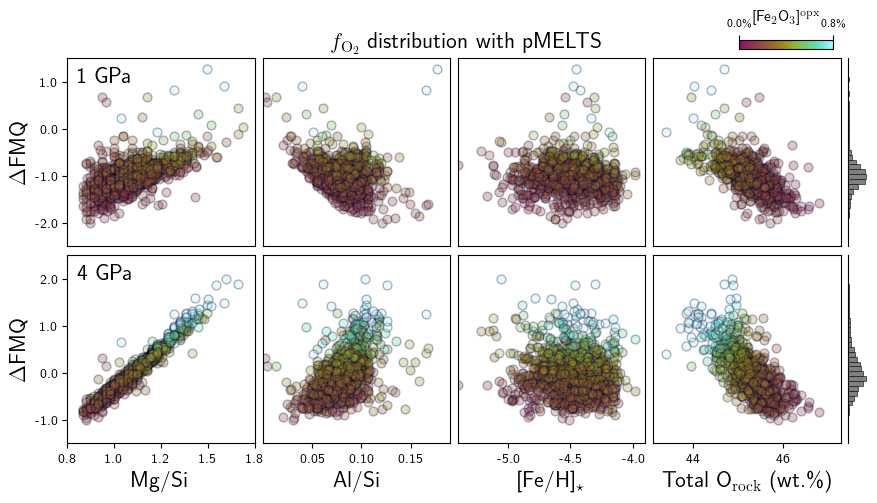}
\caption{\label{fig:xplots_mlts}Cross plots of log($f_{\ce{O2}}$) expressed as $\Delta$FMQ, resulting from pMELTS calculations, shown at $1\,{\rm GPa}$ \textit{(top)} and $4\,{\rm GPa}$ \textit{(bottom)} and $1373.15\,\text{K}$. From left to right, columns show the dependence of $\Delta$FMQ on bulk mantle Mg/Si, Al/Si, stellar metallicity [Fe/H]$_\star$, and the total refractory oxygen present in the mantle, O$_{\rm rock}$ (i.e., a sum over oxygen in all metal oxides). Each point ($N = 1198$) represents a bulk mantle composition (Ca-Na-Fe-Mg-Al-Si-O-Ti) inferred for a planet-hosting star in the Hypatia Catalog, assuming ${\rm Fe}^{3+}/\Sigma{\rm Fe} = 3$\% and ${\rm Fe}_{\rm mantle}/({\rm Fe}_{\rm core} + {\rm Fe}_{\rm mantle}) = 12$\%. Points are coloured by the \ce{Fe2O3} wt.\% composition of orthopyroxene (opx). Histograms of the $\Delta$FMQ distribution are projected on the $y$-axis.}
\end{figure*}

\begin{figure*}
\centering
\includegraphics[width=1\textwidth]{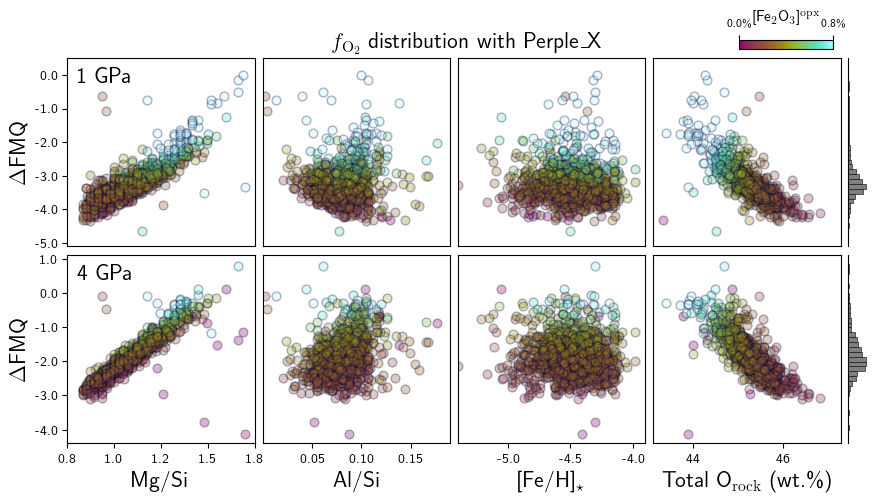}
\caption{\label{fig:xplots_px}The same results as presented in Figure \ref{fig:xplots_mlts}, but using the JH-15 database in Perple\_X (and excluding Ti from bulk compositions; $N = 1206$). Note the different $y$-axis scales.}
\end{figure*}

\subsection{The host minerals of ferric iron in planetary mantles}\label{sec:results-ferric-hosts}

JH-15 and pMELTS model the thermodynamics of ferric iron incorporation into mineral phases as particular \ferric-bearing mineral endmembers. The models differ in the endmembers they use, their solution models, and the thermodynamics of those solution models. From this fundamental description of the minerals emerges partitioning behaviour of \ferric\;(and other elements). These important differences in how the models we use are constituted leads to systematically different absolute \fo\,between them for a given bulk composition, sometimes by 2 dex (Figure \ref{fig:model_comp}). So not only is predicting \fo\,hard for a planetary mantle without prior knowledge of its ferric iron abundance, it is challenging even when we can assume its ferric iron abundance. However, we will show that between the two models we use, \textit{(i)} the amount of \fo\,variation is similar, and \textit{(ii)} the compositional dependence of the \fo\,variation is also similar. As we are focusing on \fo\,variability arising from changes in bulk mantle composition, the fact that the models agree on this suggests the ensuing insights are robust.

%\todo{Figure \ref{fig:model_comp}, not really sure how much this is adding other than illustrating perplex systematically gives lower fo2 compared to pmelts, seemingly because in pmelts ferric iron can partition into spinel. no spinel at 4 Gpa so difference is smaller here.}

The main phenomenon causing \fo\,to vary for constant bulk mantle \xfer\;is that different minerals partition characteristically different amounts of ferric iron into their crystal structures. We illustrate this phenomenon with a series of ternary diagrams (Figure \ref{fig:ferric_ternary}), showing \ce{Fe2O3} modality for each bulk composition inferred from our Hypatia sample, to better understand the origins of the \fo\,variations produced by pMELTS and JH-15. Ternary diagrams are useful for plotting three variables which sum to unity, as is the case here. 

At $1\,\text{GPa}$, the stable phases that incorporate \ferric\,are orthopyroxene, clinopyroxene, and spinel (note an olivine phase is almost always present yet contains no \ferric). Each point in the top row of Figure \ref{fig:ferric_ternary} shows the mass fraction of \ce{Fe2O3} in those phases for a given mantle composition, weighted by the mass fraction of the phase itself in the system. The fact that most points plot near the bottom-right corner shows that orthopyroxene is the most important \ferric\,host in almost all compositions at $1\,\text{GPa}$, for both models. Although spinel's crystal structure may incorporate higher fractions of \ce{Fe^3+}, when spinel is stable, its overall abundance rests at a few percent, and so has limited scope to affect the local \fo. In fact, the JH-15 model produces some mineralogies with very small spinel abundances, so many points in the top right panel of Figure \ref{fig:ferric_ternary} plot very close to the right edge of the diagram ($\sim$0\% spinel axis).

At $4\,\text{GPa}$---bottom row of Figure \ref{fig:ferric_ternary}---spinel has been replaced by garnet, which has different \ferric\,partitioning behaviour. Further, unlike spinel, garnet can start to form sizeable proportions of the mantle at a few GPa, especially for the more Al-rich planetary compositions. In the JH-15 model, orthopyroxene exchanges \ferric\,with garnet (although not shown, more \ferric\,partitions into garnet with increasing $P$). Meanwhile, the pMELTS model does not place \ferric\,into garnet, hence all points in the bottom left panel of Figure \ref{fig:ferric_ternary} plot on the $\sim$0\% garnet axis. %\todo{[which is more realistic, do we know? Oli: There is ferric iron in garnet.  So in this sense the JH15 model is more realistic]} %There is also a secondary effect of Al3+ exchange between gt and px, which substitutes with Fe3+ \citep{stolper_effects_2020}.  

Note that in order to reproduce the Earth's present-day upper mantle (``depleted mantle'') \fo\,near to FMQ, the calculations in Figure \ref{fig:model_comp} included \ce{Cr2O3} in the bulk compositions. The presence of the minor cation \ce{Cr^{3+}} can significantly influence \ce{Fe^{3+}} partitioning (even without affecting modal mineralogy; see section \ref{sec:methods-bulkcomp}). However, all subsequent calculations omit \ce{Cr2O3} because \textit{(i)} Cr is moderately volatile (hence less-related to the stellar abundance), and \textit{(ii)} Cr measurements are less-frequently listed in the Hypatia stellar compositions ($N$ = 636 in the Figure \ref{fig:model_comp} sample). The deceivingly large effect of minor cations on \fo\,highlights the fact that absolute \fo, along with being model-dependent, cannot be constrained precisely without detailed compositional information. For example, the depleted mantle composition in Figure \ref{fig:model_comp} would be 1 log unit more oxidised without Cr in the pMELTS model at $1\,\text{GPa}$.
% at 4 GPa, adding Cr lowers logfo2 by 1 dex (melts). at 1 GPa adding Cr lowers logfo2 by 0.3 dex. perplex ~0.1 change

%\todo{[todo: compare fo2 distributions with and without Cr and make sure they are the same.]} 
% for Perplex: with Cr and without has same distribution at 1 GPa, from sigma=0.51 to 0.61 (smaller with Cr) at 4 GPa. medians are similar
% for melts: with Cr has smaller sigma = 0.33 vs 0.46 at 4 GPa but only 300 runs . at 1 GPa melts smaller at 0.23 vs 0.32. Cr decreases median dQFM by -1 

%\todo{[check if there is an obvious change in phase proportions with and without Cr, or just Fe3+ partitioning---just Fe3+ partitioning]}

%%%

\subsection{Compositional correlations of \fo}

Given that the majority of \ce{Fe2O3} is in orthopyroxene for most planetary mantle compositions (Figure \ref{fig:model_comp}) and the proportion of \ce{Fe2O3} in olivine is always zero, it follows that the main predictor of upper mantle \fo\,for a given \xfer\,would be the olivine/orthopyroxene ratio. A greater olivine/orthopyroxene ratio concentrates the mantle's \ferric\,budget into orthopyroxene, thus raising the activity $a_{\rm Fe_2O_3}^{\rm opx}$ (compare (\ref{eq:K_iw})). The olivine/orthopyroxene ratio will be itself strongly positively correlated with Mg/Si \citep{hinkel_star_2018, spaargaren_plausible_2022, guimond_mantle_2023}, a fact of these phases' chemical stoichiometries: the Mg-endmember of olivine \ce{Mg2SiO4} (forsterite) requires 2 atoms of Mg per atom Si, while the Mg-endmember of orthopyroxene \ce{MgSiO3} (enstatite) requires 1 atom Mg per atom Si. Thus, excess Mg with respect to Si builds olivine-dominated mineralogies. Magnesium and silicon have this dominant role in setting planetary mineralogy simply because of their high cosmochemical abundance.

Figures \ref{fig:xplots_mlts} and \ref{fig:xplots_px} show the relationship between \fo\,(as $\Delta$FMQ) and Mg/Si, plus various other parameters describing bulk planet composition, for pMELTS and JH-15 respectively. A key result is that despite the differences between the two models (section \ref{sec:methods-mineralogy}), they nonetheless predict fundamentally similar systematics between \fo\,and planetary parameters. We emphasise that \xfer\;and thus the real \fo\,is unconstrained for exoplanets; we fix \xfer\;at 3\% for the sake of creating these figures. The Mg/Si effect that we have explained in the previous paragraph induces $\sim$3 orders of magnitude variation in mantle \fo\,(left-most columns). The scatter in \fo\,versus Mg/Si at $1\,\text{GPa}$ is wider due to slightly more complicated phase equilibria at the lower pressures; pMELTS further places more \ferric\,into spinel, with more scope for affecting \fo\,(Figure \ref{fig:ferric_ternary}). We find that the correlation of $\Delta$FMQ with Mg/Si at $1\,{\rm GPa}$ has a slope of 1.4 and 3.0, in pMELTS and JH-15 respectively, and at $4\,{\rm GPa}$, 3.4 and 4.0 (Supplementary Figure S1).

The relationship between Al/Si and \fo\,is more complex (Figures \ref{fig:xplots_mlts} and \ref{fig:xplots_px}, second columns), but again consistent between models. In pMELTS and JH-15 the correlation with \fo\,is negative in the spinel field, yet positive in the garnet field. Several effects lead to this behaviour and are discussed more thoroughly in \citet{stolper_effects_2020}. Most directly, Al-rich compositions stabilise mantles with higher proportions of the aluminous phases spinel and garnet, and less pyroxene (Supplementary Figure S2). In pMELTS at high pressure, where garnet does not take any \ferric, the \ce{Fe2O3} activity in orthopyroxene therefore increases with increasing Al/Si. In JH-15, \ferric\,is redistributed between garnet and orthopyroxene, so the trend is less straightforward and the correlation between \fo\,and Al/Si becomes weaker. Similarly at $1\,\text{GPa}$, larger proportions of spinel dilute the \ferric-bearing component in this aluminous phase, which can have a bearing on \fo.

% but not JH-15, increased garnet proportions take \ferric\,away from orthopyroxene, decreasing \ce{Fe2O3} activity in orthopyroxene, yet increasing \ce{Fe2O3} activity in garnet.... 
% \todo{opposite at 1 GPa for spinel where higher Al means more spinel which means lower ferric iron activity in Opx but spinel is such a low abundance that it doesn't contribute to fo2? [q: why does spinel not affect fo2 calcualted if it's just the activity if ferric iron in the phase that counts..?]}

The third columns of Figures \ref{fig:xplots_mlts} and \ref{fig:xplots_px} demonstrate how no correlation exists between upper mantle \fo\,and stellar Fe abundance (metallicity, [Fe/H]$_\ast$), at a fixed molar distribution of planetary iron between \ce{Fe^0}/\ferrous/\ferric. Increasing the total Fe in planetary mantles does not significantly affect the partitioning of ferric iron between minerals in pMELTS and JH-15; the ratios of \ce{Fe2O3} and \ce{FeO} activities in host minerals are largely independent of total Fe.

Lastly, in the fourth columns, we show the correlation between \fo\,and the total mass percent of refractory oxygen bound as metal oxides (e.g., O in \ce{SiO2}, \ce{CaO} etc.), which we call the total O$_{\rm rock}$ (wt.\%). Both models show that total O$_{\rm rock}$ is strongly negatively correlated with mantle \fo\,at both pressures. This seemingly counter-intuitive negative correlation is a stoichiometric effect aliasing the Mg/Si effect above; it arises from the valence state of the metal oxides forming the planet's mantle. As Mg occurs as MgO and Si as \ce{SiO2}, higher Mg/Si favours forsterite (\ce{Mg2SiO4}), which is 45.5\% O by mass, over enstatite (\ce{MgSiO3}), which is 47.8\% O by mass. Because planets with higher Mg/Si will have higher ratios of forsterite to enstatite, they will therefore have lower total O$_{\rm rock}$ for stoichiometric reasons, whilst having higher \fo\,for thermodynamic reasons. Further discussion of the Mg/Si ratio's effect on the amount of refractory O that condenses to form planets can be found in \citet{unterborn_effects_2017}.

%Note that in these figures, $\Delta$FMQ is systematically offset from what it would be if a full Bulk Silicate Earth complement of trace elements were included (see section \ref{sec:methods-components}). The values here should not be interpreted too literally; it is their variability that is important.

% This relationship between Mg/Si and refractory O is explored in detail in \citet{unterborn_effects_2017}.

% The potential effects of garnet modality in the JH-15 model are also partly affected by Mg/Si---with low-\ce{SiO2} compositions associated with greater garnet modality---but are very strongly correlated with Al content. Overall, we find that \fo\,tends to increase with garnet modality \todo{but this may just be a side effect of ol/opx...  low Mg/Si can have lots of Fe3+ in garnet --- not directly required that this lowers activity of Fe3+ because effects of both garnet abundance and Fe3+ partitioning. pMELTS shows more scatter at 1 GPa because more hosts (spinel) compared to perplex, but opposite at 4 GPa}

% \todo{[make connection to stellar nucleosynthesis setting these element ratios? --- why only Mg and Si really matter because they just have more scope for variability.]}

\subsection{Minimum variability of \fo\,across mantle compositions}

\begin{figure*}
\centering
  \includegraphics[width=\textwidth]{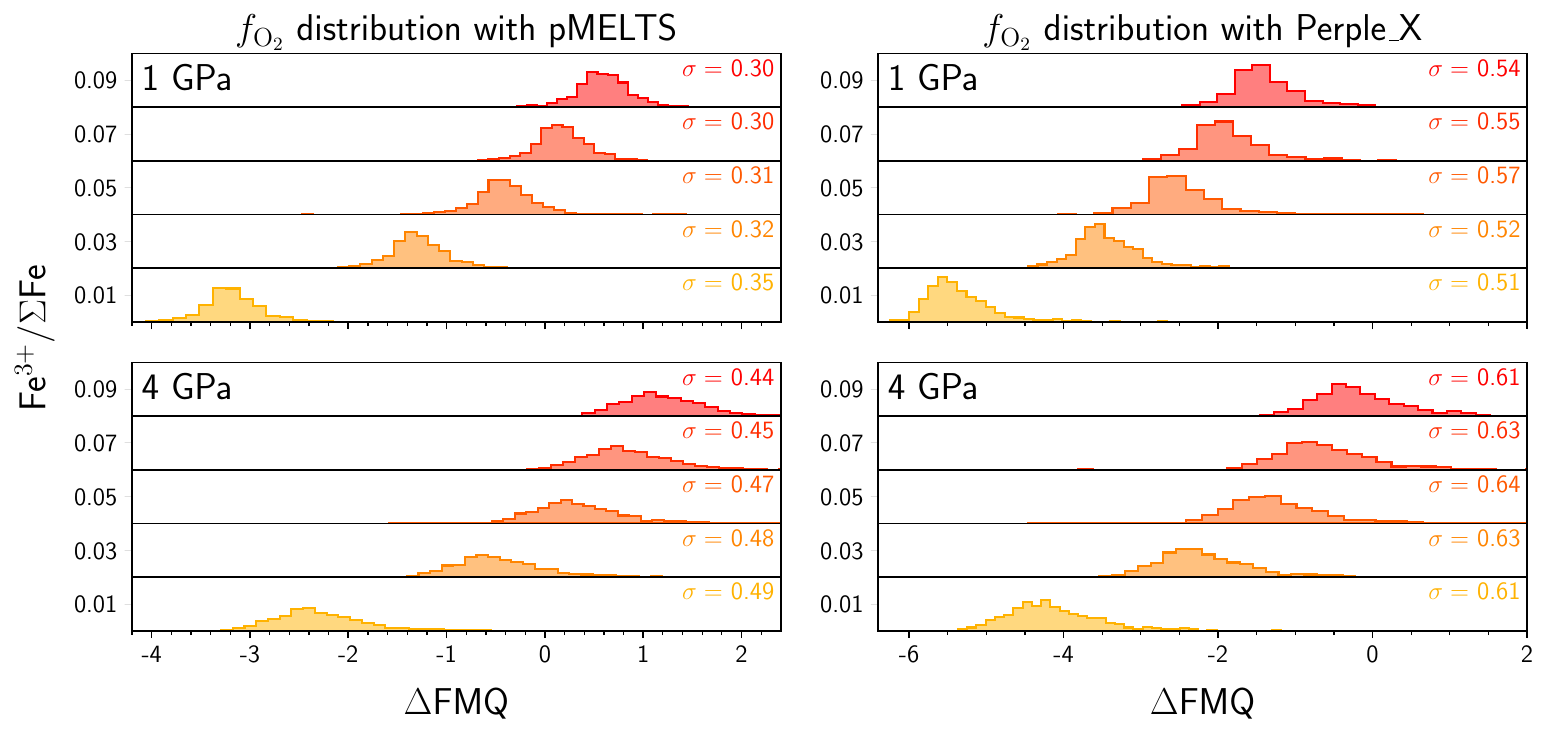}
\caption{The distributions of mantle oxygen fugacities expressed as $\Delta$FMQ, resulting from chemical variability in Hypatia host stars. Distributions are shown for different mantle \ce{Fe^{3+}}/$\Sigma$Fe ratios and assuming ${\rm Fe}_{\rm mantle}/({\rm Fe}_{\rm core} + {\rm Fe}_{\rm mantle}) = 12$\%, at $1\,{\rm GPa}$ \textit{(top)} or $4\,{\rm GPa}$ \textit{(bottom)}, using the pMELTS \textit{(left)} or JH-15 in Perple\_X \textit{(right)} models. Calculations are shown at $1473\,{\rm K}$ to ensure more compositions have numerically-stable solutions. Standard deviations $\sigma_{f\text{O}_2}$ of each distribution are shown in the upper right corners.\label{fig:hist_fe}}
\end{figure*}

\begin{figure*}
\centering
  \includegraphics[width=\textwidth]{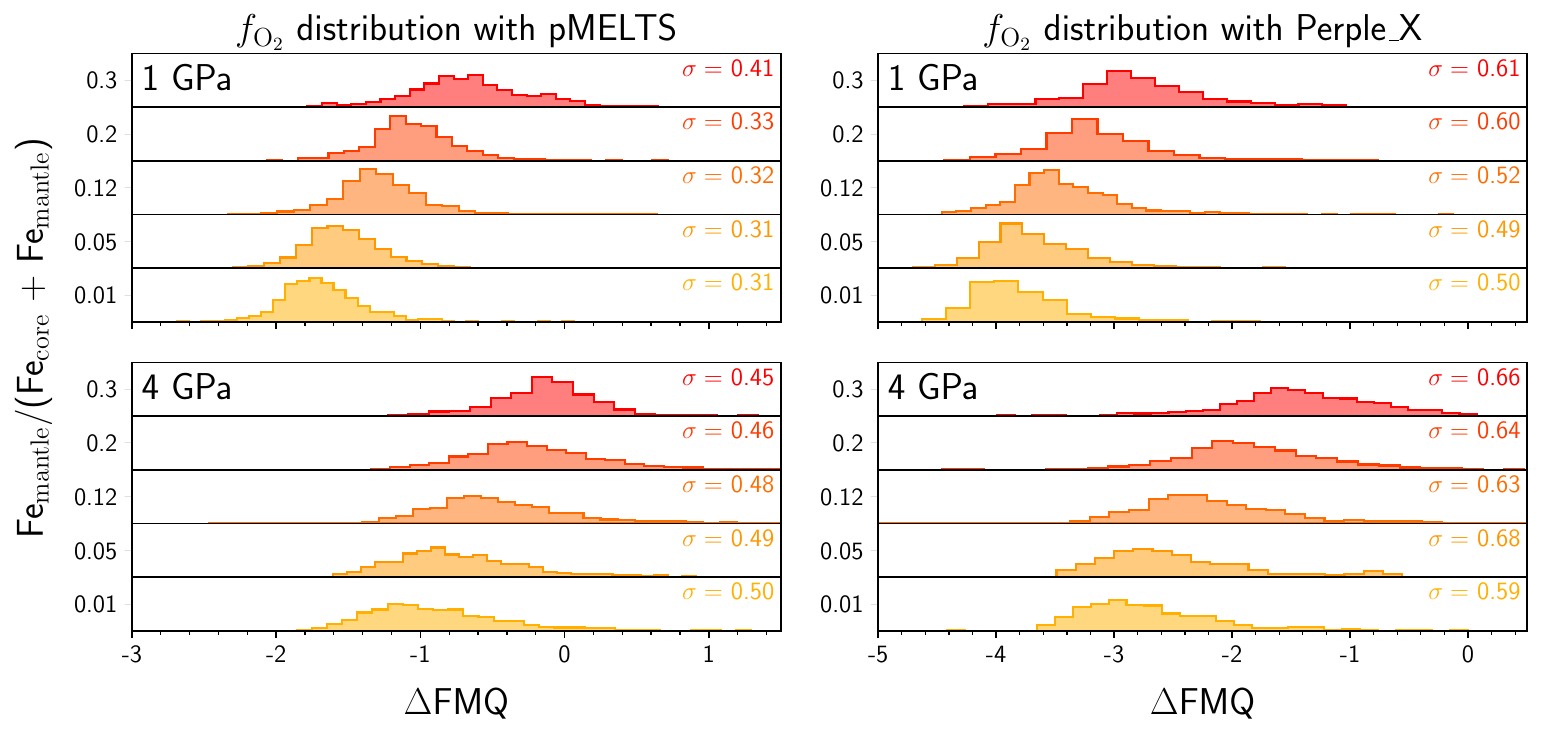}
\caption{The same as Figure \ref{fig:hist_fe}, but showing $\Delta$QFM distributions for different choices of ${\rm Fe}_{\rm mantle}/({\rm Fe}_{\rm core} + {\rm Fe}_{\rm mantle})$, assuming mantle ${\rm Fe}^{3+}/\Sigma{\rm Fe} = 3$\%. Note the different $x$-axis scales between models and between here and Figure \ref{fig:hist_fe}.\label{fig:hist_core}}
\end{figure*}

The previous subsections set up an understanding of the \fo\,variability caused by changing mineral proportions in the upper mantles of rocky exoplanets. We call our distributions a minimum variability because observations of exoplanet host star chemistry imply that these mineralogies should indeed vary across exoplanets \citep{hinkel_stellar_2014, putirka_composition_2019}. The true variability of \fo\,will be broader than what we report here if \xfer\;also varies across exoplanet mantles---empirically true, at the very least,  considering Earth and Mars \citep[e.g.,][]{dale_late_2012}. The extent of iron metal core segregation, \xcore, could also plausibly affect upper mantle \fo\,through its effect on mineralogy. We have therefore calculated exoplanet mantle \fo\,distributions as a function of these two key unknown parameters, \ce{Fe^{3+}}/$\Sigma$Fe and ${\rm Fe}_{\rm mantle}/({\rm Fe}_{\rm core} + {\rm Fe}_{\rm mantle})$ (Figures \ref{fig:hist_fe} and \ref{fig:hist_core} respectively). Here, data in each histogram encompass the same stellar abundance data for known exoplanet host stars ($N$ between 976 and 1232 for pMELTS, and between 1033 and 1251 for JH-15), but varying the fixed parameter \xfer\;or \xcore. We choose generous---yet necessarily arbitrary---ranges for these two parameters, noting that the true upper limits are unknown.

As expected, increasing \xfer\;means increasing \ce{Fe2O3} activities and therefore \fo. Figure \ref{fig:hist_fe} shows 2\%-increments of \xfer: as it increases from 1\% to 9\%, $\Delta$FMQ increases by about 4 log-units. Yet the widths of the \fo\,distributions, represented by their standard deviations, appear similar, $\sigma_{f\text{O}_2} \approx 0.3$ dex at $1\,\text{GPa}$ and $\sigma_{f\text{O}_2} \approx 0.4$ dex at $4\,\text{GPa}$. The lack of strong dependence of $\sigma_{f\text{O}_2}$ on (a reasonable range in) Fe oxidation state supports our ability to estimate a minimum \fo\,\emph{variability}. To reiterate, $\Delta$FMQ is unknown so long as we cannot constrain the ferric iron budget of a given exoplanet---the values of \ce{Fe^{3+}}/$\Sigma$Fe considered here for the purposes of calculation are somewhat arbitrary choices, and larger than the ranges considered in \citet{deng_magma_2020} and \citet{hirschmann_magma_2022} for magma oceans.

All distributions in Figure \ref{fig:hist_fe} are wider at $4\,\text{GPa}$ compared to $1\,\text{GPa}$ due to the stability of garnet, which generally has higher modality than the lower-pressure aluminous phase spinel, and thus has more scope for affecting \ferric\,activities. In detail, distributions are also slightly wider at the lowest \ferric. The non-linearity of median \fo\,versus \xfer\;is due to the fact that $\Delta$FMQ is logarithmic.%We attribute this effect to the log-scale of $\Delta$FMQ. \todo{[confirm with unlog fo2, also check phase stability?] ---an effect also seen in .}

In Figure \ref{fig:hist_core}, each histogram now represents a different choice of $\chi_{\rm Fe}^{\rm mantle}$~=~\xcore. Whilst \xfer~affects \ferric\,activities directly, $\chi_{\rm Fe}^{\rm mantle}$ merely changes the wt.\% FeO$^*$ of the mantle. Note that no compositions stabilise a Fe-metal phase in the upper mantle, so only FeO and \ce{Fe2O3} activities are relevant for \fo\,(cf. \eqref{eq:K_iw}: in the much-more-reducing magma ocean, $\chi_{\rm Fe}^{\rm mantle}$ is actively linked to \fo\,via equilibria between Fe$^0$ and Fe$^{2+}$). 

Increasing total FeO$^*$ will slightly increase the olivine/orthopyroxene ratio, but otherwise has only a passive influence on \fo. Figure \ref{fig:hist_core} shows that greater $\chi_{\rm Fe}^{\rm mantle}$ means slightly greater \fo\,at constant \xfer, which is explained by this stabilising effect of FeO$^*$ on olivine, therefore concentrating the \ferric\,in orthopyroxene (clinopyroxene, spinel, and garnet proportions are unaffected). Again, $\sigma_{f\text{O}_2}$ is similar across the large range of $\chi_{\rm Fe}^{\rm mantle}$; these distributions are reliable estimates of minimum \fo\,variabilities. The slight row-wise differences in $\sigma_{f\text{O}_2}$ do not follow the same pattern between pressures and thermodynamic datasets, which points to the complex effects of FeO$^*$ on phase stability and resulting \ferric\,activities.% and hence on \ce{FeO} and \ce{Fe2O3} activities.

%As we increase $\chi_{\rm Fe}^{\rm mantle}$, slight differences in $\sigma_{f\text{O}_2}$ have the opposite sense of change between 1 and 4 GPa and between thermodynamic datasets.%The effect of \xcore\,on median \fo\,could be amplified if we consider that higher Mg/Si in the protoplanetary disk means more refractory O is condensed and mantles are richer in FeO$^*$ for the same stellar O abundance \citep{unterborn_effects_2017}.

\subsubsection{Relevance of the core mass fraction for upper mantle \fo}

A final observation from the experiments in Figures \ref{fig:hist_fe} and \ref{fig:hist_core} is the relative importance of \xfer\;and $\chi_{\rm Fe}^{\rm mantle}$. Doubling \xfer\;from an Earth-like value increases the median $\Delta$FMQ by more than one log-unit; the same $\Delta$FMQ increase would require a much greater change in $\chi_{\rm Fe}^{\rm mantle}$. If we compare Figure \ref{fig:hist_core} with Figures \ref{fig:xplots_mlts} and \ref{fig:xplots_px}, it is clear that the direct effect of stellar abundance variability on exoplanets' upper mantle $\Delta$FMQ is more significant than the direct effect of the planets' unknown extent of core formation within the range we consider.

That is, the core mass fraction of a planet is not necessarily a proxy for its \textit{upper mantle} \fo. The equilibrium between Fe-metal and FeO that governs core size is generally no longer operating in the evolved uppermost mantle, if Fe-metal is not present here to buffer electron transfer. The direct relevance of a larger core mass fraction is just to take away from the total mantle iron budget. There are of course indirect effects (beyond the olivine stability mentioned above) in that the physical processes controlling core size may themselves impact the mantle \xfer\;(that is, $\chi_{\rm Fe}^{\rm mantle}$ and \xfer\;are probably not independent of one another in reality). For example: \textit{(i)} larger cores come at the expense of high-pressure silicate phases in the lower mantle, phases which may be important for producing \ferric\;in planetary interiors; and \textit{(ii)} certain processes that may affect the mantle \ferric\;budget would be more effective at oxidising the mantle when the mantle is more iron-poor, such as \ce{H2O} photolysis followed by \ce{H2} loss. These effects are discussed in section \ref{sec:discussion-ferriciron}.

% simulated planets across our range of \xcore\;show  \fo\,distributions shifted by only $\sim 1$ log unit. Given that varying \xcore\,from 0.01 to 0.3 as we have done encompasses a very wide range of metal partitioning,

%In our model, the total oxygen budget of the mantle can vary either by changing the ratio of core \ce{Fe^0} to the total bulk Fe inherited from the star, or by changing the ratio of \ce{Fe2O3} to total mantle \ce{FeO*}.

%\todo{nevertheless, core formation is generally not expected to suck down a planet's entire Fe budget. in particular, redox reactions with Si at core formation conditions will add some \ce{Fe^{2+}} to the silicate planet, even more-so for larger planets (fischer 2015, wordsworth 2018).}

\section{Discussion}
\label{sec:discussion}

\begin{figure*}
\centering
\includegraphics[width=0.95\textwidth]{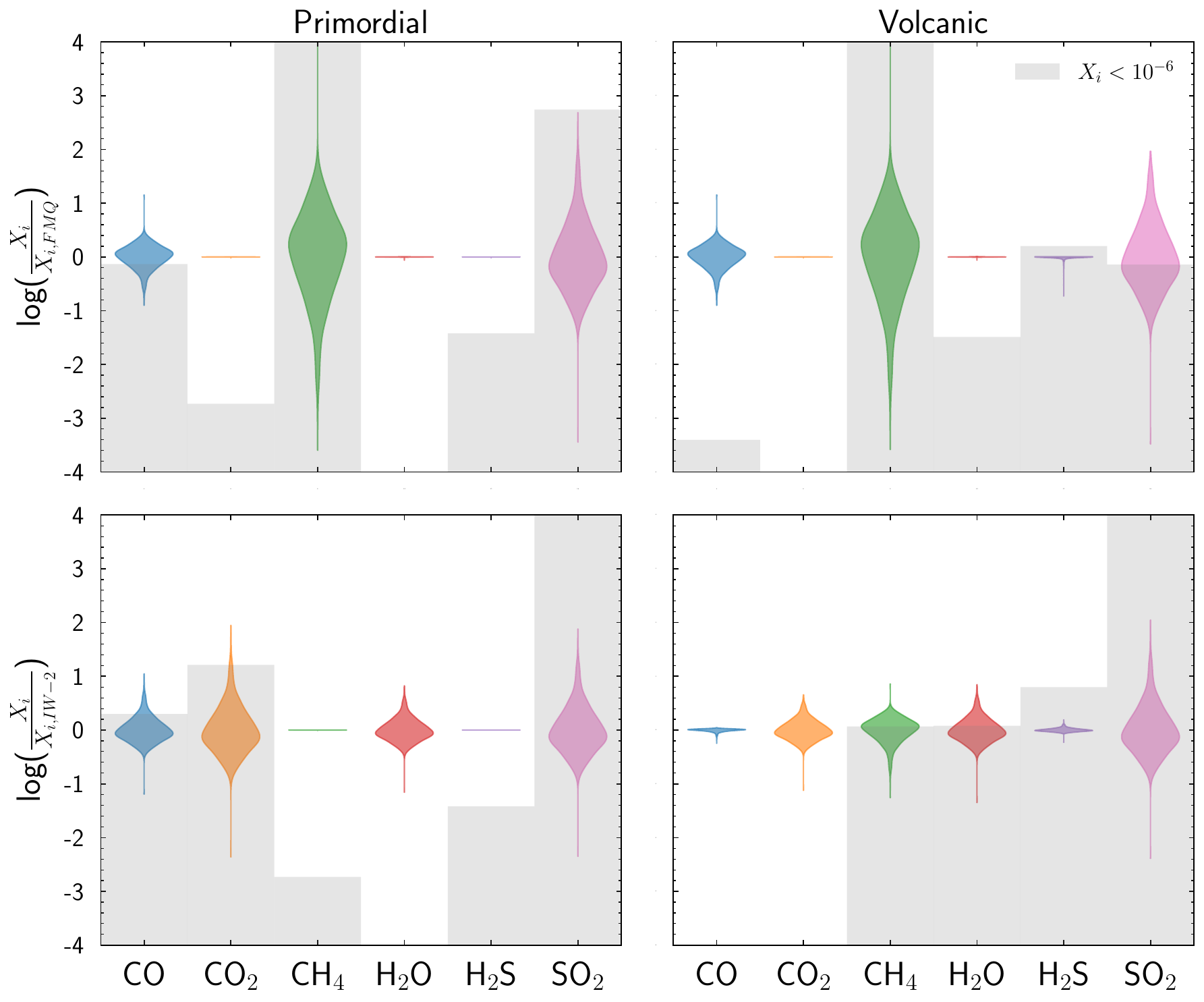}
\caption{\label{fig:speciation}Violin plots of the variation in volcanic gas speciation resulting from our calculated mantle \fo\,variability. Speciation is expressed on the $y$-axis as the the mixing ratio $X$ at the calculated \fo, relative to what the mixing ratio would be at a reference \fo---either FMQ (modern Earth's upper mantle; \textit{top}) or ${\rm IW} - 2$ (Earth's mantle at core formation; \textit{bottom})---with distributions normalised such that the median \fo\,is that of the reference. Distributions are shown for two nominal bulk atmosphere C-S-H ratios: a primordial composition \textit{(left)} reflecting the protoplanetary disk, and a volcanic composition \textit{(right)} reflecting Venus' present-day atmosphere. The grey patches show where mixing ratios are less than 1 ppm and therefore likely below the detection limit in an exoplanet atmosphere. This example calculation uses our \fo\,distribution from pMELTS at 4 GPa (nominally representative of melting below a stagnant lid), projected to a degassing temperature and pressure of 800 K and 1 bar.}
\end{figure*}

\subsection{The effect of mantle \fo\,variability on volcanic gas composition and detectability}

We choose to focus on one implication of the minimum mantle \fo\,variability, the composition of volcanic gas. Light volatile species such as \ce{CO2} and CO, \ce{H2O} and \ce{H2}, and \ce{SO2} and \ce{H2S} are carried to a planet's surface by magmas, potentially building up volcanic atmospheres \citep{holland_model_1962, gaillard_diverse_2021}. The speciation of this volcanic gas is controlled by redox reactions and hence \fo. For example, the two simple equilibria,
\begin{align}
    \ce{H_2 + 1/2O2 &<=> H_2O},\label{eq:h-gas}\\
   \ce{CO + 1/2O2 &<=> CO_2},\label{eq:c-gas}
\end{align}
will control the relative proportions of \ce{H2} and \ce{H2O} and of CO and \ce{CO2}: increasing the availability of \ce{O2} will produce more \ce{H2O} and \ce{CO2} at the expense of \ce{H2} and \ce{CO}. In reality, in atmospheres as in solid mantles, many such equilibria will be operating simutaneously, and the ultimate equilibrium proportions of gas species emerge after considering entire chemical networks.

It is an old paradigm that more-oxidising mantles produce more-oxidising volcanic gas (i.e., the right-hand sides of (\ref{eq:h-gas}--\ref{eq:c-gas})), and more-reducing mantles produce more-reducing volcanic gas \citep[and references therein]{kasting_mantle_1993}. Magmas plus any gases that exsolve from them would have their net oxidation state and \fo\,set in equilibriation with their (upper) mantle source region where melting occurs. %Here, solid-state FeO-\ce{Fe2O3} equilibria are the dominant buffer of the melt \fo\,during (low degrees of) partial melting because the mass of the mantle-melt system is dominated by the solid. 
In detail, once the melt is removed from (no longer buffered by) the mantle, degassing itself may alter the gas oxidation state insofar as certain elements, such as S or C, exchange valence electrons as they transition from the melt phase to the gas phase \citep[e.g.,][]{mathez_influence_1984, burgisser_redox_2007}, although the overall oxidising or reducing power supplied to the surface is unchanged. In the following calculations, we nevertheless make the simplifying assumption that the \textit{range} of \fo\,for a gas phase in equilibrium with a melt phase is approximately the same as the (minimum) \textit{range} of \fo\,for upper mantles which we have calculated in section \ref{sec:results}.

\subsubsection{Calculation of outgassed mixing ratios}

As an indication of how intrinsic variability in a planet's upper mantle \fo\,may impact its atmospheric chemistry, we use FastChem \citep{kitzmann_fastchem_2018, stock_fastchem_2022} to calculate gas phase mixing ratios (the number of moles of a gas phase per total moles of air) as a function of \fo. FastChem uses Gibbs free energy minimisation to obtain the equilibrium speciation of a gas from input elemental ratios. In this investigation, we focus on gases containing volatile \ce{C-O-S-H} species only. 

The \fo\,of a mixture of gas without any condensate is defined simply by the \ce{O2} mixing ratio. The input elemental oxygen abundance that will produce a \ce{C-O-S-H} gas mixture with the desired absolute \fo\,value at a given pressure and temperature is not known \textit{a priori}, but must be found by varying the input elemental abundances. Keeping all elemental abundances fixed and varying only the relative \ce{O} abundance, we obtain the elemental composition that produces a gas mixture at a desired \fo. For our choice of input elemental abundances, we choose two sets of values: \textit{(i)} solar \ce{C-S-H} ratios, to represent a primordial gas composition; and \textit{(ii)} Venusian atmospheric \ce{C-S-H} ratios, to represent a volcanic gas composition. We then run FastChem through a linear bisection algorithm that varies the relative elemental \ce{O} abundance for each set of input elemental ratios until the target \fo\,is output. This determines the gas speciation of volatiles of interest in two atmospheric types, a primordial atmosphere and a volcanically-derived atmosphere, where in both cases the redox chemistry has been set by upper mantle \fo. 

How much the mixing ratio of a species will change with \fo\,depends on where the system is located in \fo-space. For example, in a simple C-O system \eqref{eq:c-gas}, the shift from CO-dominated to \ce{CO2}-dominated speciation occurs around ${\rm \Delta FMQ} - 2$; above this point there is little subsequent variation in $X_{\ce{CO2}}$. Therefore, in addition to the two scenarios of elemental abundances, we also consider two nominal scenarios for where the mantle \fo\,distribution is centred: \textit{(i)} at ${\rm IW} - 2$, representing Earth's early reduced mantle at the time of core formation; and \textit{(ii)} at FMQ, representing Earth's evolved, oxidised upper mantle. 

To obtain a volcanic gas mixing ratio distribution, we re-centre the mean of the pMELTS \fo\,distribution (Figure \ref{fig:xplots_mlts}) on FMQ and on ${\rm IW} - 2$, to obtain two sets of \fo distributions each for the primordial and volcanic compositions. Re-centering is approximately equivalent to changing the wt.\% composition of \ce{O2} or the mantle \xfer. We use the \fo\,distribution at $4\,\text{GPa}$ to reflect the mantle source under a thick stagnant lid \citep[see section \ref{sec:methods-pressures}]{noack_volcanism_2017, gaillard_diverse_2021}, representing a prototypical rocky planet \citep{stern_stagnant_2018}. For each \fo\,value in the distribution, we run FastChem in $0$\,D, at $1$\,bar and $800$\,K to obtain the equilibrium composition of a hypothetical gas outgassed at the planet's surface. We emphasise that these calculations express the equilibrium mixing ratios of the gas \textit{input} to the atmosphere for constant \ce{C-O-S-H} gas composition; we do not model further processing in the atmosphere such as photochemistry, equilibriation with an ocean or crust, etc., nor do we model the solubility of these species in the melt. 

Figure \ref{fig:speciation} shows the resulting variation in the abundances of the volcanic gases \ce{CO, CO2, CH4, H2O, H2S} and \ce{SO2}, which can be up to tenfold due essentially to mantle Mg/Si (i.e., the first-order driver of mineralogy in the magma source region). This figure can demonstrate when the mantle \fo\,minimum variability affects the composition and detectability of volcanic gas species in exoplanet atmospheres. In many cases, all mixing ratios remain below a nominal $1\,\text{ppm}$ detection limit, but there are mentionable exceptions.

For carbon species, mineralogical \fo\,modulation has order-of-magnitude effects on \ce{CO2} and \ce{CH4} mixing ratios, and can modulate \ce{CH4} from undectable to detectable in reduced volcanic atmospheres---notable, given \ce{CH4}'s possible service as a biosignature gas \citep[e.g.,][]{wogan_abundant_2020}. Detectable outgassed CO mixing ratios can change by several-fold due to mantle Mg/Si variations, even with a moderately oxidising Fe redox state in the upper mantle. 

For sulphur species, there is clearly a large effect of mantle Mg/Si on the directly-outgassed mixing ratio of \ce{SO2} in all four scenarios, all else equal. This effect may be enough to raise outgassed \ce{SO2} above a nominal detection threshold in an oxidised volcanic atmosphere. Potentially large changes in directly-outgassed \ce{SO2} mixing ratios, even at very trace, undetectable absolute amounts, may still be consequential for photochemistry and aerosol formation, and their associated effects \citep[e.g.,][]{loftus_sulfate_2019, jordan_photochemistry_2021}.

Finally, Figure \ref{fig:speciation} shows that nominally-detectable amounts of \ce{H2O} might be outgassed from an upper mantle with a ``reduced'' Fe redox state, for the upper half of the Mg/Si distribution. 

Whilst the calculations in Figure \ref{fig:speciation} were performed at a nominal $1\,{\rm bar}$ and $800\,{\rm K}$, different pressures or temperatures affect the volcanic gas chemistry. Pressure variations from 0.1 to 10 bar generally have minor effects on the shapes of the mixing ratio distributions, but do affect the overall abundance of certain species and so their detectability, namely CO and \ce{CO2} in primordial atmospheres and \ce{CH4} and \ce{H2O} in volcanic atmospheres (Supplementary Figures S3--S4), in addition to pressure's solubility effects mentioned below. The trends with temperature are more variable, and species that straddle our potentially observable boundary switch with changing temperature from 600 to 1000 K (Supplementary Figures S5--S6). More detailed studies of exoplanet outgassing would treat atmospheric pressure and temperature sensitively.

\subsubsection{Other consequences of \fo\,on mantle outgassing: mantle-melt partitioning and gas solubility}

Some oxidised species appear to be much more soluble in basaltic melts than reduced counterparts; for example, \ce{H2O} vs. \ce{H2}, and \ce{CO2} vs. CO and \ce{CH4} \citep[and references therein]{gaillard_theoretical_2014}. At $\sim$1 bar we can assume most of thsee volatiles do degas, but at Venus-like atmospheric pressures (= magma degassing pressures), the higher solubilities of these oxidised species will limit their tendency to degas from the magma. Further, the fraction of S that degasses from a basaltic melt even at 1 bar has a strong positive correlation with \fo\,below FMQ, which would amplify the \ce{SO2} variability in Figure \ref{fig:speciation} \citep[and references therein]{gaillard_redox_2015}. These phenomena point to a general effect of \fo\,on the total mass of volatiles outgassed---depending on the atmospheric pressure---which compounds the effect of \fo\,on gas phase speciation.

In a similar phenomenon, the movement of carbon from the solid mantle into partial melt may be limited at \fo\,below IW, where carbon takes the form of graphite \citep{holloway_highpressure_1992, holloway_graphitemelt_1998, grott_volcanic_2011, wetzel_degassing_2013, armstrong_speciation_2015, li_carbon_2017}.  Therefore, at modest atmospheric pressures, the less-efficient extraction of carbon from such reduced mantles could limit the total mass of carbon species they can outgas, all else equal \citep{guimond_low_2021}.

\subsection{Some reasons for planet-planet \xfer\;variability}\label{sec:discussion-ferriciron}

Whilst we have calculated the distributions of $\Delta$FMQ across a na\"ive range of fixed \xfer\;values (Figure \ref{fig:hist_fe}), the true underlying distribution of \xfer\;is unknown. Below are several ways whereby \ferric\;could be produced in planetary interiors, each implying different upper limits on \xfer:

\begin{enumerate}
\item \textit{Endogenous processes}
\begin{enumerate}

    \item \textit{Iron disproportionation---}The crystallisation of perovskite from the magma ocean triggers a reaction where FeO disproportionates into Fe-metal plus \ce{Fe2O3} (see section \ref{sec:quantifying-how-oxidising}). \ce{Fe2O3} stabilises perovskite, whilst Fe-metal likely sinks to the growing core. This process, sometimes called mantle self-oxidation, can occur on planets large enough to reach perovskite stability pressures \citep[e.g.,][]{frost_experimental_2004, wade_core_2005, wood_accretion_2006, frost_redox_2008}. Hence self-oxidation (or lack thereof) has been put forward as a reason why Mars' mantle appears more reducing than Earth's---Mars' mantle is too shallow to stabilise perovskite \citep{wade_core_2005}. The amount of \ferric\,that can be produced by disproportionation is hard-limited by the total amount of Fe in perovskite \citep[plus postperovskite;][]{catalli_xray_2010}. Deeper mantles would produce more \ferric, which may then be slowly homogenised in the mantle through solid-state convection. However, if Fe-metal \textit{does not} sink to the core, which may be the case on the most massive rocky planets due to their turbulent magma oceans \citep{lichtenberg_redox_2021}, then there is not necessarily a continuous trend between the size of a planet and how oxidised its mantle is.
    
    \item \textit{Pressure-dependent \ferric\,stability in the magma ocean---}Higher pressures increasingly favour \ferric\,over \ce{Fe^{2+}} in molten silicates, due to the smaller partial molar volume of the former \citep{kress_compressibility_1991, hirschmann_magma_2012, armstrong_deep_2019, deng_magma_2020, kuwahara_hadean_2023}. This process is distinct from (i.a) in that it can occur in the magma ocean itself before it crystallises (so \ferric\,is more rapidly homogenised). Fe-metal still must be lost to the core in order to effectively oxidise the magma ocean. If the mantle inherits the resulting \xfer\;of the magma as it crystallises, then deeper magma oceans will also make mantles with higher \xfer, due to this \ferric\,stability effect. In this way, magma ocean depth may similarly be a principal determinor of overall mantle \xfer\;\citep{deng_magma_2020}, with the same caveat as in (i.a) of needing Fe-metal to sink.
     
    \item \textit{Cr oxidation during magma ocean crystallisation---}Because Cr is present as CrO in the deep magma ocean, but as \ce{Cr2O3} in the solid mantle, CrO may be oxidised during magma ocean crystallisation via the reaction \ce{2CrO + Fe2O3 = Cr2O3 + 2FeO}, which could decrease \xfer\;significantly (by a factor of two for Earth) depending on the accreted Cr budget \citep{hirschmann_magma_2022}. This process represents one reason why the mantle \xfer\;could differ from the magma ocean value. 
    \end{enumerate}
    
\item \textit{Exogenous processes}
\begin{enumerate}

    \item \textit{Oxidation by volatile accretion}---A substantial layer of volatiles such as \ce{H2O} above a magma ocean can be a powerful source of oxidation potential. In this case, the upper limit to how much \ferric\,is produced from Fe$^{2+}$ is essentially unconstrained, given unknown initial abundances of volatiles. For example, \ce{H2O} dissociation in the atmosphere followed by \ce{H2} escape to space produces \ce{O2}, which can then be absorbed by the magma ocean to oxidise FeO into \ce{Fe2O3}. \citep{schaefer_predictions_2016}. Similarly, \ce{H2O} dissociation in a magma ocean produces \ce{H2} and \ce{Fe2O3}; the \ce{H2} degasses, increasing \xfer\;of the magma ocean \citep{sharp_hydrogenbased_2013, sharp_nebular_2017}. After the magma ocean crystallises, the early crust could also be a sink for atmospheric free oxygen---which, if subducted or otherwise buried, could also contribute to mantle oxidation \citep[e.g.,][]{krissansen-totton_oxygen_2021}. Late accretion of \ferric -rich chondritic material has also been proposed to have oxidised Earth's mantle  \citep{oneill_origin_1991}. Meanwhile, metal-rich enstatite chondritic material or even organic C in carbonaceous chondritic material could plausibly cancel out this oxiding power \citep{hirschmann_magma_2022}.
    
   \end{enumerate} 
\end{enumerate}

% FeO-Fe2O3 partitioning in melt
% Cr equilibria producing Fe2O3

\subsection{The underlying rocky planet Mg/Si distribution?}\label{sec:discussion-mgsi}

Mantle compositions will only represent host star refractory element abundances if no additional processes partition these elements relative to one another whilst planets form and differentiate. However, several processes may increase Mg/Si relative to the primordial protoplanetary disk \citep{ringwood_significance_1989, tronnes_core_2019, miyazaki_dynamic_2020}. A salient example is that the Sun's composition produces an Mg/Si ratio of 1.05, yet measurements of Earth mantle xenoliths suggest Mg/Si $\approx$ 1.24 \citep[the bulk mantle may not be well-represented by these upper mantle xenoliths;][]{javoy_integral_1995}. A small systematic offset in bulk planet Mg/Si would manifest as a small translation of the \fo\,distribution. Since our model merely captures the effect of Mg/Si variation on \fo, it is agnostic to what actually causes this variation (e.g., stellar abundances alone, or stellar abundances plus secondary processes). Because it seems unlikely that secondary processes would converge to \textit{decrease} the amount of inter-planet Mg/Si variation, our reported \fo\,variability remains a minimum. Although our results are therefore not strictly sensitive to the true median of the Mg/Si distribution of rocky planets, it is worth briefly discussing the degree to which a planet's Mg/Si might be modulated away from its host star. Because mantles with Mg/Si approximately $\in [0.7, 1.5]$ will have pyrolitic (i.e., pyroxene- and olivine-dominated) compositions \citep{guimond_mantle_2023}, a median Mg/Si offset of at least $-0.3$ or $+0.5$ is necessary to significantly alter the mineralogically-derived spread in \fo.

Some Si likely enters the metal core \citep[e.g.,][]{ringwood_chemical_1959, javoy_integral_1995, wood_accretion_2006, schaefer_metalsilicate_2017}. On an Earth mass planet, partitioning Si into a metal core such that the core is 15 wt.\% Si, 32.5\% of the planet mass, and contains 88\% of the planet's iron atoms would increase a ``solar-derived'' mantle Mg/Si to 1.42, for example. We choose to not explicitly model metal/silicate Si partitioning---i.e., Fe is the only element modified by planetary differentiation---in order to avoid unconstrained and unnecessary complexity. In particular, the value of the metal/silicate partitioning coefficient depends on the temperature and pressure conditions at the base of the magma ocean.  %, and in any case we exclude the few \todo{(how many?)} compositions that produce stable quartz or coesite \ce{SiO2} phases. 
However, if Si decreasingly partitions into metal at higher pressures \citep{schaefer_metalsilicate_2017}, then more massive planets may evolve upper mantles which are systematically slightly pyroxene-richer, and thus more reduced. 

\subsubsection{Oxygen fugacity of silica-saturated mantles}\label{sec:discussion-sio2}

Although we excluded compositions from our results which stabilise pure \ce{SiO2} phases, these mineralogies could exist at mantle ${\rm Mg/Si} \lesssim 0.7$ (or $\sim$6\% of the Hypatia stars in our sample), depending on the mantle FeO$^*$ content. %This situation becomes more likely if Si core partitioning is inefficient; e.g., on massive rocky planets \citep{schaefer_metal-silicate_2017}. 
Silica-saturated mineralogies have a distinct pattern of \fo: when they exist, we find a composition-independent \fo, consistently about 2 log-units higher than the median for the given \xfer. %These planets could be speculatively more likely to outgas \ce{O2} directly, given sufficiently high \xfer\;\citep{wordsworth_redox_2018}. 
However, the thermodynamic data is not well-constrained here, and more experiments are needed to understand these hypothetical exotic mantle mineralogies.
% 75 cases with SiO2 at Earth-like core partitioning

% \subsection{WD mg/si?}

% \subsection{Effects of minor oxides on \fo}\label{sec:discussion-minor-oxides}

% \todo{i.e., depending on the solution model, the amount of minor oxides such as \ce{Cr2O3} and \ce{TiO2} can have non-negligible effects on ferric iron partitioning in the upper mantle, even if they do not have a large effect on the overall mantle mineralogy... this points to the difficulty in deterministically predicting the upper mantle \fo\,for an unknown exoplanet, as \ce{Cr} in particular is moderately volatile and hard to track from star to planet \citep[see][]{wang_detailed_2022}.}

% \todo{check if there is an obvious change in phase proportions with and without Cr, or just Fe3+ partitioning.}

\subsection{Limitations due to thermodynamic data}

In working with the two models JH-15 and pMELTS, our results naturally inherit their limitations. In particular, the thermodynamic activities determined by these datasets and solution models are not neccessarily applicable to the wide range of exoplanet compositions that we consider here, and the associated caveats apply to our results. As an example, the models assume olivine is \ferric -free under all conditions. Even in the compositional ranges for which they were developed, the two models contain different assumptions: Perple\_X includes \ferric\;in its garnet solution model, whilst pMELTS does not (Figure \ref{fig:ferric_ternary}). 

On the basis of their similar spreads of \fo (despite dissimilarities like that with the garnet model), we have nevertheless argued that a minimum \fo\,variabilty is robust. Importantly, both models include ferric iron in pyroxene: this is a key reservoir for determining how oxidising a planet's source of magmatism is. However, the partitioning of \ferric\,between pyroxene and melts is poorly constrained experimentally \citep[although there have been recent attempts;][]{rudra_experimental_2021}. Thus future models, informed by new experiments, may differ particularly in how mantle \fo\,evolves during progressive melting. In sum, we anticipate that more experiments in the near future, based in experimental petrology or first-principles molecular dynamics, would extend the reliable thermodynamic data into the wider---yet likely finite---compositional space of rocky planet mantles.

%However, caveats still apply to our results. For example, we are assuming that olivine is always \ferric -free. This appears to be essentially the case for olivine in natural peridotites. Small amounts (a tenth of the pyroxene \xfer\;in the same rock) have been measured in mantle xenoliths \citep[e.g.,][]{ejima_oxidation_2018}. Whilst this trace amount still implies very low \ferric\;activity in olivine and thus would not have a large effect on \fo, the Fe$^{3+}$/Fe$^{3+}$ partitioning in mineral assemblages with especially low-pyroxene modalities is not well-constrained. 

\section{Conclusions}
\label{sec:conclusion}

This work has identified a tractable component of the \fo\,distribution across exoplanet mantles: we have calculated the minimum width of the distributions due to mineral phase equilibria at 1 and $4\,\text{GPa}$. The actual centre (and to a lesser extent the true width) of the distribution remains unknown unless we know what ratios of Fe$^{3+}$ to Fe$^{2+}$ to expect for these planets. Nevertheless, we find similar \fo\,standard deviations for widely variable values of mantle \xfer, pointing to a robust minimum \fo\,variability that ultimately reflects the measured refractory element distribution across \textgreater 1000 host stars from the Hypatia Catalog. 

Whilst \xfer\;might be said to set the bulk oxidation state of the upper mantle---Fe is by far the most abundant multivalent rock-forming element---the relationship between \xfer\;and \fo\,is not unique because the various mineral equilibria controlling \fo\,will also shift with bulk composition \citep{frost_introduction_1991}. We have demonstrated this fact by finding a wide range of absolute \fo\,for planets with identical \xfer\;but plausibly different proportions of olivine, pyroxene, and spinel or garnet.

In particular, we show that a planetary mantle’s Mg/Si ratio alone has an order-of-magnitude effect on the \fo\,of its upper mantle. This effect is comparable to the effect on upper mantle \fo\,of decreasing FeO$^*$ by locking Fe metal in the core. Mg/Si controls the relative proportions of the abundant minerals olivine and pyroxenes; since pyroxenes can incorporate \ferric\,in their crystal structures but olivine cannot, increasing pyroxene proportions with decreasing Mg/Si will dilute (lower the activity of) the \ferric-bearing component in pyroxenes, hence lowering the \fo\,of the system \citep{stolper_effects_2020}. This effect induces compositional variability in mantle \fo\,across rocky planets.

To illustrate one planetary consequence of this mantle \fo\,variability, we predict the distributions in volcanic gas composition that directly result from our \fo\,distribution, given that the \fo\,of these mantle-derived gases inherits the \fo\,of the mantle source region \citep[e.g.,][]{gaillard_redox_2015}. Depending on the C-H-S ratio and median \fo, the resulting mixing ratios of outgassed CO, \ce{CH4}, and \ce{SO2} can differ by up to ten-fold, due to mantle mineralogy variations alone. Overall, our results establish another role for mantle composition to play in the detectability of atmospheric species on rocky worlds and in quantifying abiotic baselines for interpreting biosignatures.

\section*{Acknowledgements}

We acknowledge the support of the University of Cambridge Harding Distinguished Postgraduate Scholars Programme and the Natural Sciences and Engineering Research Council of Canada (NSERC). Cette recherche a \'et\'e financ\'ee par le Conseil de recherches en sciences naturelles et eng\'enie du Canada (CRSNG). The research shown here acknowledges use of the Hypatia Catalog Database, an online compilation of stellar abundance data as described in Hinkel et al. (2014, AJ, 148, 54), which was supported by NASA's Nexus for Exoplanet System Science (NExSS) research coordination network and the Vanderbilt Initiative in Data-Intensive Astrophysics (VIDA). This manuscript has benefitted from a thoughtful review by Jie Deng, as well as from discussions with Amy Bonsor and Sami Mikhail.

\section*{Data Availability}

The Python code and the pMELTS and Perple\_X input files used in this study are available from the authors upon request.

\bibliographystyle{mnras}
\bibliography{references.bib}

\end{document}